\documentclass{elsart}
\usepackage{epsfig}
\usepackage{natbib}
\usepackage{amssymb}
\journal{Physica A}

\begin{document}

\begin{frontmatter}

\title{ A self-adjusted Monte Carlo 
        simulation \\
        as a model 
        for financial markets with central regulation}

\author{Denis Horv\'ath \corauthref{cor}}, 
\corauth[cor]{Corresponding author.}
\ead{horvath.denis@gmail.com} 
\author{Martin Gmitra, Zolt\'an Kuscsik }
\address{Institute of Physics, \v{S}af\'arik University, 
Park Angelinum 9, 040 01 Ko\v{s}ice,\\ Slovak Republic}

\begin{abstract}
Properties of the self-adjusted Monte Carlo algorithm applied to 2d~Ising 
ferromagnet are studied numerically. 
The endogenous feedback form expressed in terms of the instant 
running averages is suggested in order to generate a biased random walk 
of the temperature that converges to criticality without 
an external tuning. The robustness of a stationary regime with 
respect to partial accessibility of the information is demonstrated. 
Several statistical and scaling aspects have been identified 
which allow to establish an alternative spin 
lattice model of the financial market. 
It turns out that our model alike model suggested by
S.~Bornholdt, Int.~J.~Mod.~Phys.~C {\bf 12} (2001) 667,
may be described by L\'evy-type stationary distribution 
of feedback variations with unique exponent $\alpha_1 \sim 3.3$.
However, the differences reflected by Hurst exponents 
suggest that resemblances between the studied models seem 
to be nontrivial.
\end{abstract}


\end{frontmatter}

\section{Introduction}

During the past decades, the financial markets around 
the world have become more and more interconnected. 
The financial globalization has changed the organization 
structure of the stock markets by creating 
new risks and challenges for market participants. 
Dramatic expansion of the cross-border 
financial flows as well as domestic flows within 
the countries due to very rapid increase of 
telecommunication and computer-based products
have emerged. As a consequence, 
the world financial markets are increasingly
efficient today than ever before. 

The integration and globalization process opens discussion about 
the activity of the central economic institutions that focus attention 
to financial market monitoring. The subsequent regulatory
legislative yields asymptotically to dynamic balance 
between coalitions based on agreements of the cooperating 
companies and on the other hand competitors 
that enhance a diversity of the price trends. 
The {\em feedback} signals of different reliability are 
monitored and daily evaluated by central economic 
institutions. Consequently, this affects correlated decisions  
of the market participants. An expected effect 
of the global self-regulation is a portfolio 
efficiency.

Recently, there have been many attempts to gain a control over 
the dynamics of complex systems. The approaches based on the principles 
of feedback and adaptivity emerges as a perspective branch~\cite{Melby2000}. 
The importance of the positive feedback mechanism in the economic context 
has been studied in Ref.~\cite{YoussHolyst2000}.
An approach based on the feedback control of spatially extended optical
system has been employed in the Ref.~\cite{Bleich1996}. The feedback 
mechanism is also exploited in Monte Carlo~(MC) algorithm introduced 
by us recently in Ref.~\cite{HorvathGmitra2004}. The algorithm is formulated 
as a random walk of temperature variable biased by feedback that mixes 
current and past stochastic signals, originating at the platform 
of extended statistical system. The time integrated signals are used 
to build up the actions shifting an instant temperature towards the 
critical temperature value. Due to limited memory depth, 
the convergence process is necessarily accompanied 
by uncertainty. Since the feedback treats a partial information only, 
many parallels with risky decisions within the business 
world can be found. Therefore, the MC feedback model may be used
for simulation of decision-making in the stochastic environment. 

The plan of the paper is as follows. In the next section the 
self-adjusted MC algorithm is reintroduced briefly. 
Its relationship to non-equilibrium spin updates is discussed in~Sec.~\ref{Sca1}.
The~Sec.~\ref{SOCrel1} deals with the statistics of spin clusters near 
the critical conditions and the self-adjusted algorithm is related to 
cellular automaton models of self-organized criticality~(SOC). 
In~Sec.~\ref{ECO11}, the applications of mentioned MC dynamics to models 
of financial markets is presented. 

\section{The self-adjusted random walk near criticality}

Before proceeding to the introduction 
of financial model, we recall the 
salient points of the self-adjusted 
MC model~\cite{HorvathGmitra2004} that
is based on the random walk of the 
temperature variable $T_t$ defined 
by the one-step recurrence  
\begin{equation} 
T_{t+1} = T_t +  \xi_t \Delta\, 
\mbox{sign}(F_t)\,,
\label{Ttttt1}
\end{equation}
where $\Delta$ parameterizes a maximum step length $|T_{t+1} - T_t|$,
randomized by $\xi_t$ that is uniformly 
distributed within the $\langle -1,1\rangle$ range. 
The subscript $t$ accounts for 
MC steps per $N$ degrees of freedom. 
The {\em feedback regulatory response}
\begin{equation}
F_t = 
\frac{
   \langle  E^3  \rangle_{t}  - 
3  \langle  E    \rangle_{t}
   \langle  E^2  \rangle_{t}
 + 2 \langle E \rangle_{t}^3}{  T_t^4 N }
- 2 \frac{ \langle E^2 \rangle_{t} - \langle  E \rangle_{t}^2
}{T_t^3 N}\,,
\label{DCDT1}
\end{equation}
mixes an influence of bias and randomness 
involved in the energy series $\{E_t\}$ that represents 
outcome of some extended statistical system. 
Eq.(\ref{DCDT1}) approximates a temperature 
derivative of the specific heat  
$\frac{\partial}{\partial T} [
(\,\langle E^2 \rangle_{\mathrm{ eq}}-\langle E\rangle_{\mathrm{ eq}}^2\,
)/(T^2 \,N)]$
at the canonical equilibrium~(eq) 
ensemble defined by the constant 
temperature $T \sim T_t$ \cite{Binder1988}. 
In Eq.(\ref{DCDT1}) the canonical averages are 
approximated by instantaneous running averages 
of the energy cumulants 
$\langle E^p\rangle_{t}$,\, $p=1,2$ 
calculated with the assistance of  
linear filter~\cite{Principe2000} 
as it follows
\begin{equation}
\langle E^p \rangle_{t} =  
(1 - \eta) \langle E^p \rangle_{t-1}  
+ \eta E_t^p \,\,.
\label{Eqppp1}
\end{equation}
Here the plasticity parameter
$\eta\in (0,1)$
reweights an 
energy samples $E_t$ 
according to delay.  
For $\eta\rightarrow 0$, running 
averages vary slowly 
and long-memory processes dominates, 
whereas in the limit 
$\eta \rightarrow 1^{-}$ 
the memorizing 
becomes 
faster but 
less efficient.

\begin{figure}
\begin{center}
\epsfig{file=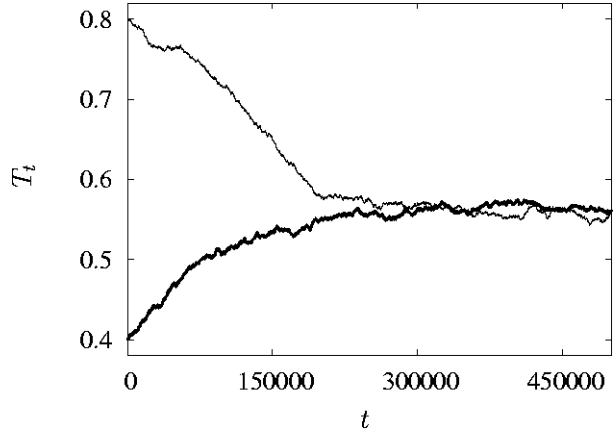,width=8.0cm}  
\end{center}
\caption{
Temporal dependence of self-adjusted MC model for
transient $T_t$ dynamics obtained for $L=90$ and 
parameters $\eta=0.01$, $\Delta=10^{-5}$. Two paths [starting from 
$T_{t=0}=0.4$, and $0.8$] tend to the unique 
attractor $T_{\rm c}(L)$ (asymptotic regime), as the 
variable $t$ increases.}
\label{fIG1}  
\end{figure}

The feedback is applied to maintain 
the critical regime of the 2d~Ising spin micromodel of ferromagnet. 
This is defined for 
$i=1,2, \ldots, N=L^2$ sites of 
$L\times L$ square lattice
characterized by the exchange energy
$E = -  \sum_{\mbox{\tiny nn}}  S_i S_j$,
$S_i \in \{-\frac{1}{2}, \frac{1}{2}\}$,
where exchange coupling constant 
rescales an instant temperature $T_t$.
The nearest neighbor~(nn) interaction 
is considered. This interaction picture 
is supplemented by the periodic boundary
conditions in all directions.
The standard Metropolis 
single-spin-flip 
algorithm~\cite{Binder1988,Metropolis}
where two states are linked by 
the flip acceptance probability
$p_{\mathrm{acc}} = 
\min \{1,\,\exp\left(- \delta E_i/T_t\right)\,\}$
depends on the energy change $\delta E_i$ 
at $i$th site. 
The technical issue of the 
proposed implementation is that $T_t$ is hold 
constant during $N=L^2$ spin 
updates~(i.e.~1 MC step per sample).  
The example which illustrates $T_t$ dynamics is shown in Fig.\ref{fIG1}. 
As follows from Eq.(\ref{Eqppp1}), a filtered information 
serves to provide $T_{t+1}$ near criticality.
In compliance with a previous facts,
the {\em feedback role} in dynamics can be summarized as follows:
a)~it {\em receives} series of energy $\{E_t\}$ 
values from extended statistical system
(or only sufficient part of the series is served for this purpose);
b)~it {\em treats} the series 
via the linear {\em filtering} 
that generates {\em running averages} $\langle E^p\rangle_t$; 
c)~it transmits the {\em regulatory signals} 
$\mbox{sign}(F_t)$ via 
the temperature 
given by 
Eq.(\ref{Ttttt1}). 

After the 
transient time the thermal noise overcomes a regular bias involved in stochastic $F_t$
and system with feedback maintain a stationary regime. Due to fluctuations of $T_t$ a 
non-equilibrium distribution is generated. Therefore, the expectation 
mean values, worth our interest are given by
\begin{equation}
\langle \, T^p\, \rangle = 
\frac{1}{ n_1-n_0}
\sum_{t'=n_0}^{n_1} T^p_{t'}\,\,,\qquad p=1,2\,.  
\end{equation}
where $n_1\gg n_0$.  
If the number of excluded steps 
$n_0$ is larger 
than the transient time, 
the average $\langle T \rangle$  
estimates critical point of the extended system. 
Thus, the observation time is defined by $\tau_{\mathrm{obs}}=n_1-n_0$. 
In our simulations 
we used 
$n_0\simeq 10^5$, 
$n_1\simeq 10^7$.

\section{The finite-size effects of $\langle T\rangle$}\label{Sca1}

There are many examples of the simple non-equilibrium spin-flip 
stochastic models~\cite{Buonsante2002,Garrido87} giving rise to 
the various dynamical phases. 
To simulate nearly equilibrium systems, the
specific non-equilibrium approaches have been suggested.
For instance, the Ref.~\cite{Tomita2001} exploits
the controlled growth of percolating spin cluster.

The postulation of artificial dynamics clearly 
places novel time scales into original 
equilibrium problem. In our case the combination 
of two parameters $\eta$ and $\Delta$ affect 
the distance between a stationary non-equilibrium statistics 
and canonical equilibrium.
The basic properties of stationary 
non-equilibrium regime 
can be summarized as follows:
\begin{description}

\item 1.  
The stationary regime with a ferro-paramagnetic order and 
positive $\langle T \rangle$ is stabilized by a sufficiently 
slow dynamics limited to $\Delta<\Delta_{\rm tr}(\eta)$, where
$\Delta_{\rm tr}(\eta)$ is a threshold value.
The irreversible drop to negative 
$\langle T\rangle$~(antiferromagnetic order) 
occurs at $\Delta=\Delta_{\rm tr}$. 

Similar transition can be found for the coupled map lattice
where antiferromagnetic order is generated 
spontaneously by ferromagnetic couplings accompanied by
sufficiently fast synchronous dynamics~\cite{Angelini2003}.
For $L=10$, 
$\eta=10^{-3}$ 
we estimated $\Delta_{\rm tr} 
\sim 2.2 \times 10^{-3}$.

\item 2. The canonical equilibrium 
limit $\langle T \rangle \rightarrow T_{\rm c}(L)$, 
where $T_{\rm c}(L)$ is 
the pseudocritical temperature 
of specific heat, is reached 
for the sufficiently slow dynamics $1\gg\Delta>0$,
when the feedback memory is deep ($1\gg\eta$)
and parameters admit the scale 
separation $\eta \gg \Delta$. 
In addition, the stationary averages have 
to be accumulated 
for 
$\tau_{\mathrm{obs}}\gg 1/\eta$.

\item 3. For stationary 
regime~(see point 1) the 
nonzero variance 
$\langle (T-\langle T\rangle)^2\rangle$ 
gives rise to the broad energy distribution 
in comparison to Boltzmann one.

\end{description}

\begin{figure}[h]
\begin{center}
\epsfig{file=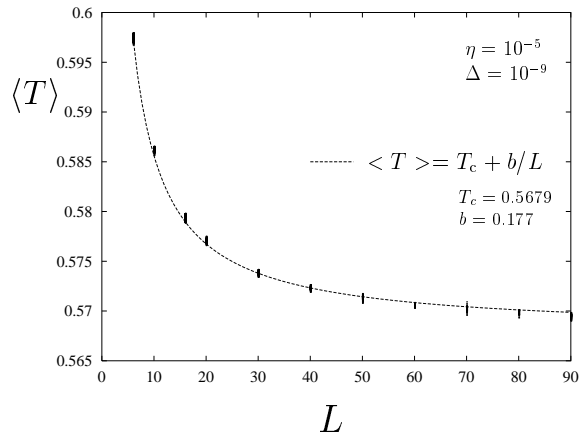,width=8.0cm}  
\end{center}
\caption{
The non-equilibrium dynamics confirms $1/L$ finite-size dependence of  
pseudo-critical temperature. True critical value $T_{\mathrm{c}}$ 
is estimated from the fit 
$\langle T \rangle  = T_{\mathrm{c}} + b/L$ 
for $L=10,20, \ldots,90$.}
\label{fIG2}  
\end{figure}

The question of practical relevance is whether true $T_{\mathrm{ c}}$ can be 
attained by non-equilibrium dynamics. The result of simulation realized 
for $\eta=10^{-5}$, $\Delta=10^{-9}$ is depicted in Fig.\ref{fIG2}. 
The standard finite-size dependence of $T_{\mathrm{c}}(L)
\simeq 
\langle T \rangle =
T_{\mathrm{c}}+ b/L$~\cite{Binder1988}
(where $b$ denotes the thermal coefficient) 
has been obtained from the fit.
Without additional tuning the simulation provides estimate 
$T_{\mathrm{c}}\simeq 0.5679$  
of the equilibrium 
exact value $T_{\mathrm{c}}^{\mathrm{ex}} \simeq \,[2\ln(1+\sqrt{2})]^{-1} 
\simeq 0.56729$~\cite{ExactSolOnsager1944}\,.

\begin{figure}
\begin{center}
\epsfig{file=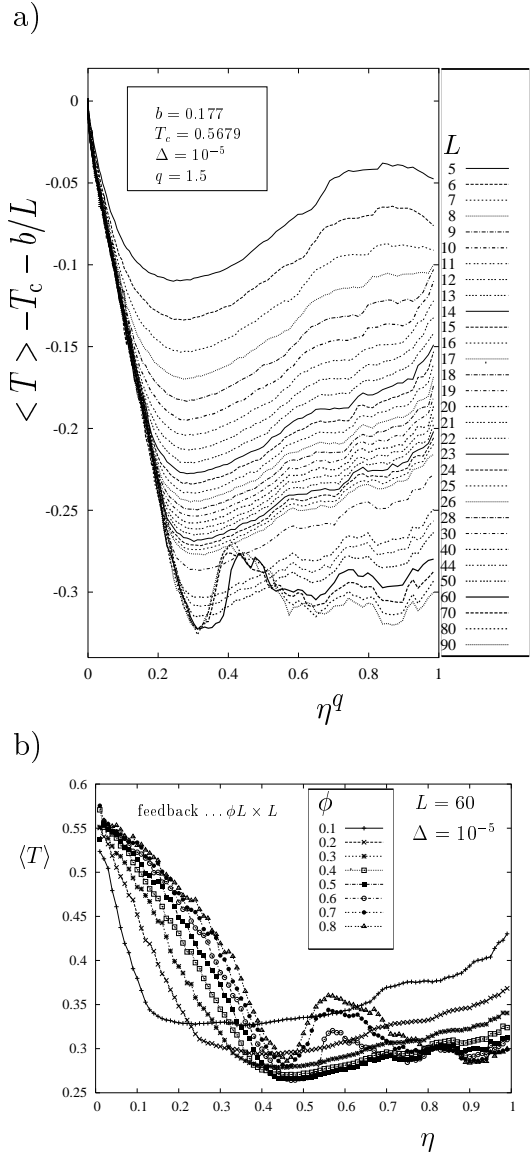,width=7.0cm}
\end{center}
\caption{  
a)~The  $\eta$-dependence of $\langle T \rangle$ calculated for
$L=5,6,7,8,9,\ldots,70,80,90$ lattices.  
The mean temperature 
decreases with 
$\eta$ for $\eta<\eta_{\mathrm{c}}(L)$. 
The upper bound is chosen 
to guarantee that
$\langle T\rangle$ 
calculated for different lattices 
merge to the unique $\eta^q$ asymptotics. 
b)~The robustness~(or failure tolerance) 
of averages with respect 
to feedback that 
accumulates partial 
statistics for the rectangular segment 
$[\phi L] \times L$, 
$0<\phi<1$ of $L\times L$ lattice.}
\label{fIG3}  
\end{figure}

\begin{figure}
\begin{center}
\epsfig{file=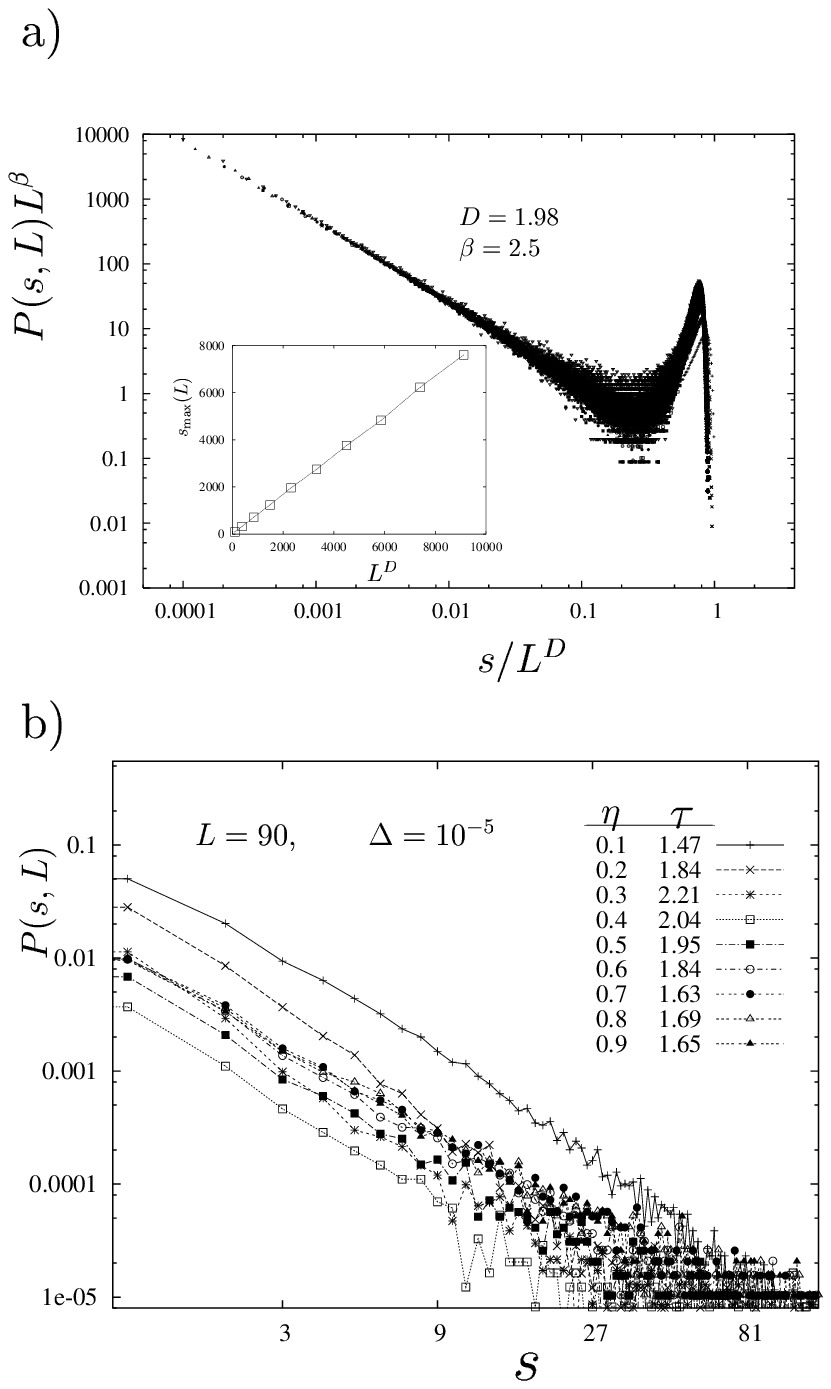,width=8.0cm}
\end{center}
\caption{  
a)~The pdf's of spin cluster for $L=10$, $\ldots,$ $100$. 
The inset shows a lattice size dependence of the peak 
position $s_{\mathrm{max}}(L)\propto L^D$. 
The simulation is carried out for parameters 
$\eta=10^{-2}, \Delta=10^{-5}$.
b)~A log-log plot of spin 
cluster size pdf 
considered for different $\eta$. 
The exponent 
$\tau$ obtained 
from the fit 
$P(s,L) \propto  s^{-\tau}$ 
within the range $s<30$. 
The differences in $\tau\in (1.47, 2.21)$ 
are connected with 
changes~$\langle T \rangle$~(see Fig.\ref{fIG3}).}
\label{fIG4} 
\end{figure}

As one can see in Fig.\ref{fIG3}a follows that substantial finite-size 
corrections occur for a short-time memory. In that case, the temperature 
mean value can be interpolated by
\begin{equation}
\langle T  \rangle= 
T_{\mathrm{c}} + \frac{b}{L} - A \eta^{q}\,, \qquad 
\Delta \ll \eta < \eta_{\mathrm{c}}(L)\,,
\end{equation}
where cut-off $\eta_{\mathrm{ c}}(L)$ 
is roughly 
interpolatated by $\eta_{\mathrm{ c}}(L)
= \eta_{\infty}-0.343/L$ with 
$\eta_{\infty}=0.468$. 
From the fit carried out within the range
$\eta \in \langle \,0,\, 
\frac{\eta_{\mathrm{ c}}}{2}\,\rangle$ 
we have determined the exponent 
$q\simeq 1.5$.  
Assuming that the lowest mean value
$\langle T\rangle \sim T_{\mathrm{ min}} \sim T_2(L)$,
where $\eta_{\infty}$ 
is achieved, for $L\rightarrow \infty$ 
we may write
$A=(T_{\mathrm{ c}} - T_{\mathrm{ min}})/\eta_{\infty}^q$. 
The presence of the temperature lower bound
is associated with the condition 
that peak of specific heat is formed. 
This implies the requirement of a residual acceptance 
of single-spin-flips 
at very low temperatures. 
The bound can be explained in 
the frame of two-level specific heat model denoted as $C_2$.
The model takes into account only single branch of  
ground state $-L^2/2$~(since the ergodicity is broken, 
the ground state is counted once) and 
$L^2$-fold single-spin excitation of energy~$2-L^2/2$. 
The condition for extreme 
${\rm d} C_2/{\rm d}T|_{T=T_2}=0$ yields 
equation $(T_2-1 )\exp(2/T_2) + L^2 ( 1+T_2 )=0$. 
The last relation predicts a monotonous 
decrease of $T_2$ with $L$ and solution can be simply approximated by
$T_2\sim \frac{1}{\ln L}$ for infinite $L$ asymptotics.
To estimate $T_2$ we have carried out 
simulation for $L=90$ and $\eta\in (0.45, 0.55)$ 
that gives $\langle T \rangle$ from $(0.25, 0.3)$ range. The mean value 
$\langle T \rangle$ is approximated by $T_2=0.212 \sim \frac{1}{\ln(90)}=0.222$.

The assumption about the energy gap corresponding to two lowest 
levels is consistent with the presence of the crossing over 
memory depth $1/\eta_{\mathrm{c}} \sim 2$~(two MC steps).
To maintain a stationary regime the substantial 
memory reduction for $\eta>\eta_{\mathrm{ c}}$ 
have to be compensated by the enlarged $\langle T\rangle$ value. 
We observed that for $\eta\sim \eta_{\mathrm{ c}}$ 
the ergodic time $\tau_{\mathrm{ e}}$ between 
the plus and minus magnetization stages diverges 
as $\tau_{\mathrm{ e}}\propto L^z$, where $z\simeq 2$ is the dynamical 
exponent~\cite{LandauBinder2000}. If $\tau_{\mathrm{ obs}}$ 
is smaller than $\tau_{\mathrm{ e}}$, 
the broken symmetry can be detected. 
Clearly, for $\eta>\eta_{\mathrm{ c}}$ the "jaggy" 
$\langle T \rangle (\eta)$ dependence indicates the occupancy 
of only few lowest energy levels.

\section{The relationship to SOC and percolation models}\label{SOCrel1}

The concept of SOC proposed 
by Bak~\cite{Bak88} 
represents the unifying theoretical 
framework of systems that drive themselves 
spontaneously to critical regime.
The sand pile, forest-fire~\cite{Drossel92} 
and game-of-life \cite{Alstrom94} 
are well-known 
examples of cellular automaton SOC models. 
Their stationary 
regime can be characterized by invariant 
{\em probability distribution functions}~(pdf's) 
of spatial or temporal 
measures of the dissipative 
events called~{\em avalanches}.  
The corresponding pdf's include a power-law 
intervals with the specific 
non-equilibrium 
critical exponents. 

According to Ref.~\cite{Kadanoff}, the stationary SOC regime implies the 
operation of inherent feedback mechanism. On the contrary, our model has 
explicitely defined global feedback coupled to the MC dynamics. 
Such coupling is completely different from the standard models of SOC, 
but the arguments given in this section suggest that additional coupling 
anchors statistical properties indistinguishable from SOC models.
 
In the next, we focus our attention to the problem of 
{\em partial information regulation} to answer the question, how 
important can be a concrete feedback form.  We consider 
spatially truncated form of feedback defined by energy cumulants calculated 
for the rectangular $[{\phi} L] \times L$ lattice segment, where 
$\phi \in \langle 0,1\rangle$ and $[x]$ denotes an integer value of $x$.
The results are depicted in Fig.\ref{fIG3}b for $L=60$.
The simulation has confirmed that stationary regime is robust
with respect to the {\em partial information regulation} represented by 
$\eta$ and $1-\phi$ terms. As one can be see in Fig.\ref{fIG3},
both sufficiently small $\eta$, ($\eta<\eta_{\mathrm{c}}(L))$, 
and $1-\phi$ have similar influence on diminishing the stationary 
$\langle T\rangle$ value.

As pointed out in Ref.~\cite{Sornette}, the pdf's of spin clusters 
at a second-order critical point resembles pdf's of the spatial extent of avalanches. 
Admittedly, the correspondence between spins and clusters is 
not unique and thus numerous mapping schemes could be examined. 
We have utilized the well-known mapping to bond percolation model~\cite{Fortuin72}
via Wolf's algorithm~\cite{Wolff89} to identify cluster without overturning.
To interpret the cluster statistics, we suggest the universal scaling in the form
\begin{equation}
P(s,L)= L^{-\beta}\, 
g\left(\frac{s}{L^D}\right)\,,
\label{scaclus1}
\end{equation}
where $s$ is the number of spins belonging to the same cluster. 
In Fig.\ref{fIG4} are depicted spin cluster pdf's. As one can see, 
the distribution functions for clusters $s \ll s_{\mathrm{ max}}(L)$ 
clearly obey the scaling given by Eq.\ref{scaclus1}, where 
$s_{\mathrm{ max}}(L)$ is proportional to the size of a spanning cluster.
It is associated with the peak $P(s_{\mathrm{ max}},L)$ in $P(s,L)$ 
distribution. Using the scaling relation $s_{\mathrm{ max}}(L)\propto L^{D}$,
the fractal dimension $D\simeq 1.98$ has been determined from the fit, see
inset in Fig.~\ref{fIG4}a. Assuming an asymptotic form $g(x)\propto x^{-\tau}$ in the limit 
$x\rightarrow 0$ we obtained $P(s,L) \propto  L^{ -\beta  + \tau D} s^{-\tau}$ that 
implies independence on $L$ if $\beta=\tau \,D$.
Surprisingly, the dependencies in Fig.\ref{fIG4}b indicate that small-scale 
power-law scenario survives for full $\eta\in (0,1)$ range. It doesn't matter
how distant an assembly is from the thermal equilibrium except a spanning 
cluster that grows with increasing $\eta$.

\begin{figure}
\begin{center}
\epsfig{file=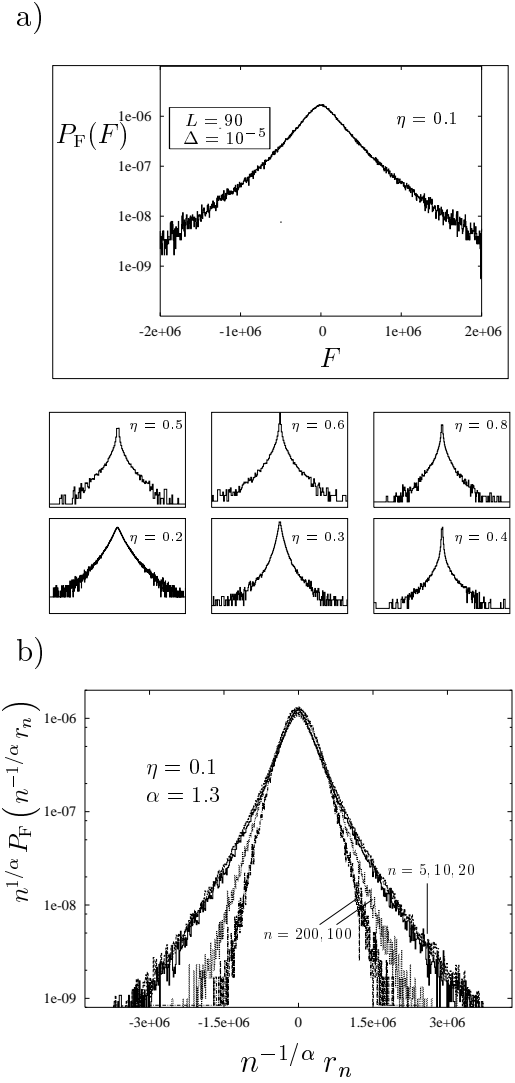,width=7.0cm}
\end{center}
\caption{  
a)~The pdf of $F_t$ shows a high leptocurticity that survive  
at different choices of $\eta$.  
b)~The scaling property given by Eq.(\ref{scalingalpha1}) and 
crossing over from the L\'evy to Gaussian pdf for $n=5,10,20$. 
The fluctuations of $r_{n,t}$, $n<20$ merge to the unique scaling 
function for $\alpha \simeq 1.3$.}
\label{fIG5} 
\end{figure}

\begin{figure}
\begin{center}
\epsfig{file=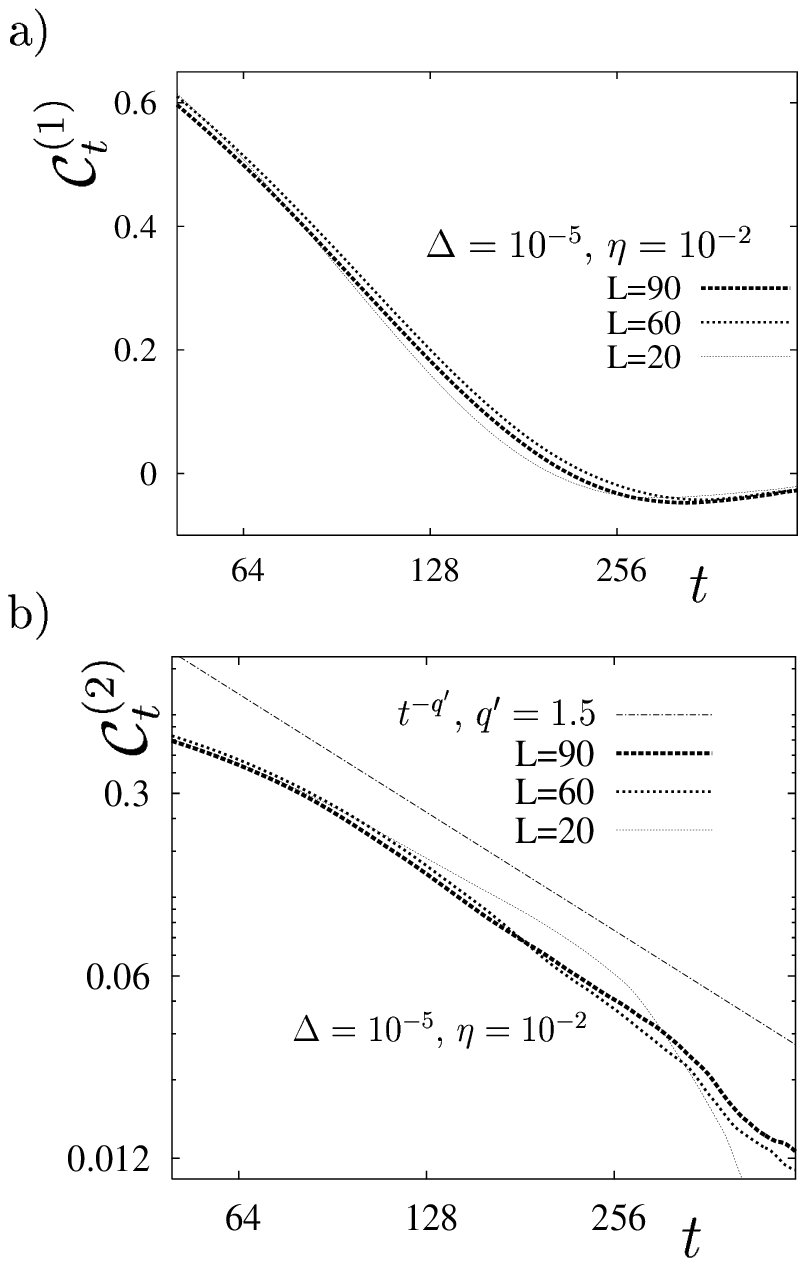,width=7.0cm}
\end{center}
\caption{ a)~Semilogarithmic logarithmic of autocorrelation function ${\rm C}^{(1)}_t$ 
at given interval demonstrate the standard exponential dependence and
the long-time anti-correlation zone is also formed.  
b)~Logarithmic plot of the autocorrelation function ${\rm C}^{(2)}_t$ 
exhibits power-law regime described by the slope $(-q')=-1.5$. 
The autocorrelation functions indicate the typical time scale of the order 
$O(1/\eta^2) = O(100)$.}
\label{fIG6}  
\end{figure}
Since the static 
bond percolation 
model pays no attention 
to temporal variations, the alternative maps should be examined, 
for instance those taking into account a short-time memory of 
cluster evolution~\cite{Mouritsen92}.
The question whether differences in a spin - cluster map definition 
might affect a non-equilibrium exponent $\beta$, introduced in Eq.\ref{scaclus1},
is left for future research.

\section{The relationship to financial market models}\label{ECO11}

\subsection{Statistical properties relevant for financial market modeling}\label{ECO12}

Performed simulations for pdf of $F_t$ defined as 
$P_{\rm F}(F)\equiv\langle\delta_{F,F_t}\rangle$ confirm that flat tails exhibit
within the full range of $\eta$, see Fig.\ref{fIG5}.
Therefore, it is worth to check {\em functional stability} of aggregated 
regulatory responses 
\begin{equation}
r_{n,t} =
\sum_{t'=t}^{t+n} F_{t'}\,.
\end{equation}
The results of analysis depicted in Fig.\ref{fIG5}b has validated the expected scaling 
\begin{equation}
P_{\mathrm F}(r_n)= 
r_n^{-1/\alpha}\, f\left(r_n n^{-1/\alpha} \right)
\label{scalingalpha1}
\end{equation}
for $n<10$ with estimated exponent $\alpha \simeq 1.3$, where
$f(\cdot)$ is some universal function. Since $F_t$ is bounded,
the crossing-over from L\'evy to Gaussian scaling has been 
verified, see Fig.\ref{fIG5}b. More profound connection to the price 
dynamics is discussed in the following section. 

The dynamics of self-adjusted system 
can be analyzed by means of autocorrelation functions 
of the time lag $t$:
\begin{eqnarray}
{\rm C}^{(p)}_t &= &  
\frac{ \langle F_{t'}^p \, 
               F_{t'+t}^p\, 
\rangle - 
\langle F_{t'}^p \rangle \langle F_{t'+t}^p \rangle
}{ 
\langle (\,F_{t'}\, )^{2p} \rangle - 
(\langle F_{t'}^p \rangle)^2  
}  \,, \qquad p=1,2\,. 
\end{eqnarray}
The MC model yields nearly exponential 
decrease of ${\rm C}^{(1)}_t$, see Fig.\ref{fIG6}. 
The power-law dependence of ${\rm C}^{(2)}$ which becomes 
more pronounced for larger $L$ is identified.
Assuming the form ${\rm C}^{(2)}_t \sim t^{-q'}$
we have determined $q'\simeq 1.5$.  

\subsection{The financial market model formulation}

On the basis of previous facts we have proposed model based on the two 
postulates: (i)~under the conditions of new economy 
the world market entities (participants, companies) 
are forced to evolve towards the longer auto-correlation times.
It means that hardly predicable - volatile price 
statistics gets stuck in the regime which 
imply highest degree of uncertainty.
It might be classified as an~{\em edge of the chaos}. 
Such interpretation is fundamentally  equivalent with the dynamics 
of the continuous phase transitions and dynamics of information
processing~\cite{Langton1990};
(ii)~the centralized institutions are established to satisfy 
the global information and legislative needs 
of market entities looking for the arbitrage opportunities. 
The action of participants leads to enhancement of market effectiveness 
in the sense of the postulate~(i).

Our simulation have shown that presented spin model 
that involves feedback has much in common 
with Ising spin market models of the price dynamics~\cite{Ponzi2000}. 
The feedback introduced in Bornholdt's model~\cite{Bornholdt2001}, which
was later elucidated by Kaizoji, Bornholdt and Fujiwara (BKF), 
is linked to the local in time and global in lattice space magnetization 
$m_t=(2/L^2) \sum_{i=1}^{L^2} S_{i,t}$. 
Both models are inspired by the 
paradigm of minority game~\cite{Challet1997}. 
Speaking in game theory term, the feedback promotes 
the continuing tournaments between competing ordered~(ferro) and 
disordered~(para) magnetic phases. To maintain a balance between 
ordered ferromagnetic and disordered - paramagnetic spin phase, 
the feedback mechanism including a global cooling or heating is applied. 
The predicted temperature is simultaneously 
transmitted to all lattice sites. Clearly, the 
effect of such step has probabilistic nature, 
because a freedom exists between the market entities which can uphold or abolish 
older or establish new coalitions, respectively. Subsequently, any decision 
is reflected by correlations between entities. 

\begin{figure}
\begin{center}
\epsfig{file=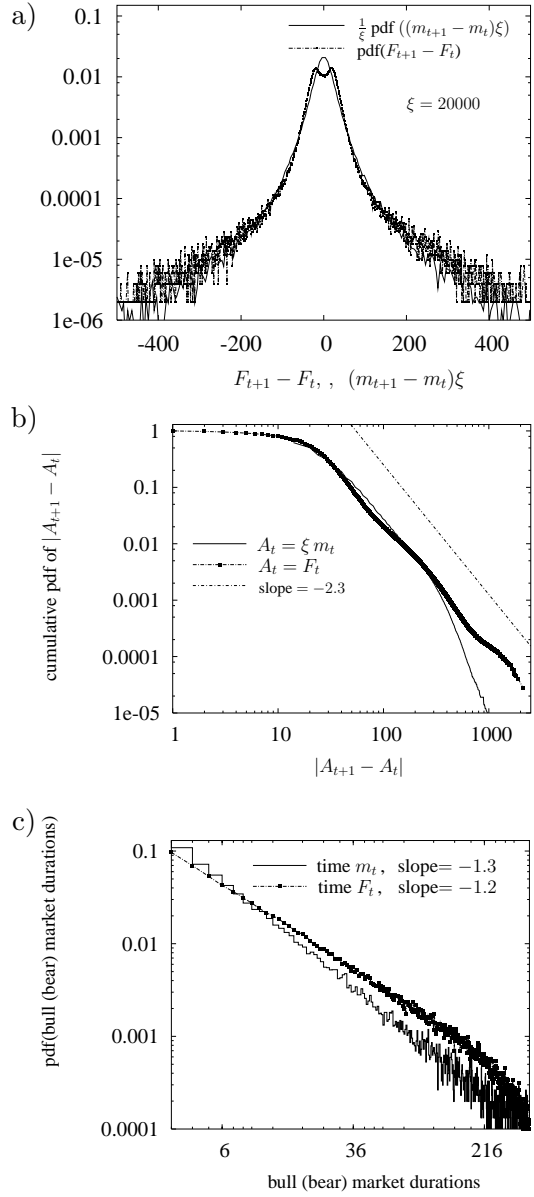, width=7.0cm}
\caption{
The comparison of BKF model (solid lines) and model introduced in this paper
for the parameters listed in the table~\ref{Tabbb1}. a)~the rescaling 
factor $\xi$ is used to attain a full data agreement except the anomaly for 
small lags $F_{t+1}-F_t$. b)~the oscillatory deviations from the power-law
has been observed. Despite of this, the rough consistency of cumulative 
volatility of returns can be attained by rescaling. 
c)~the pdf's of bull (bear) regime durations. 
}
\label{fIG7} 
\end{center}
\end{figure}

Two principal situations may be observed from the view 
point of the proposed model. At the subcritical temperatures 
the clusters grow, which means that correlations 
among market entities become increase. A subcritical 
temperature can increase a mean cluster size 
until a spanning cluster come into existence. 
The ferromagnetic domain state can be understood 
as a market situation where competition is distorted 
due to monopoly, duopoly or trust formation effects. 
To prevent from highly correlated stocks, the feedback triggers 
the antitrust operations~\cite{Wrigley1992}. 
The portfolio diversification is represented by the 
fragmentation of the spin clusters.

\begin{figure}
\begin{center}
\epsfig{file=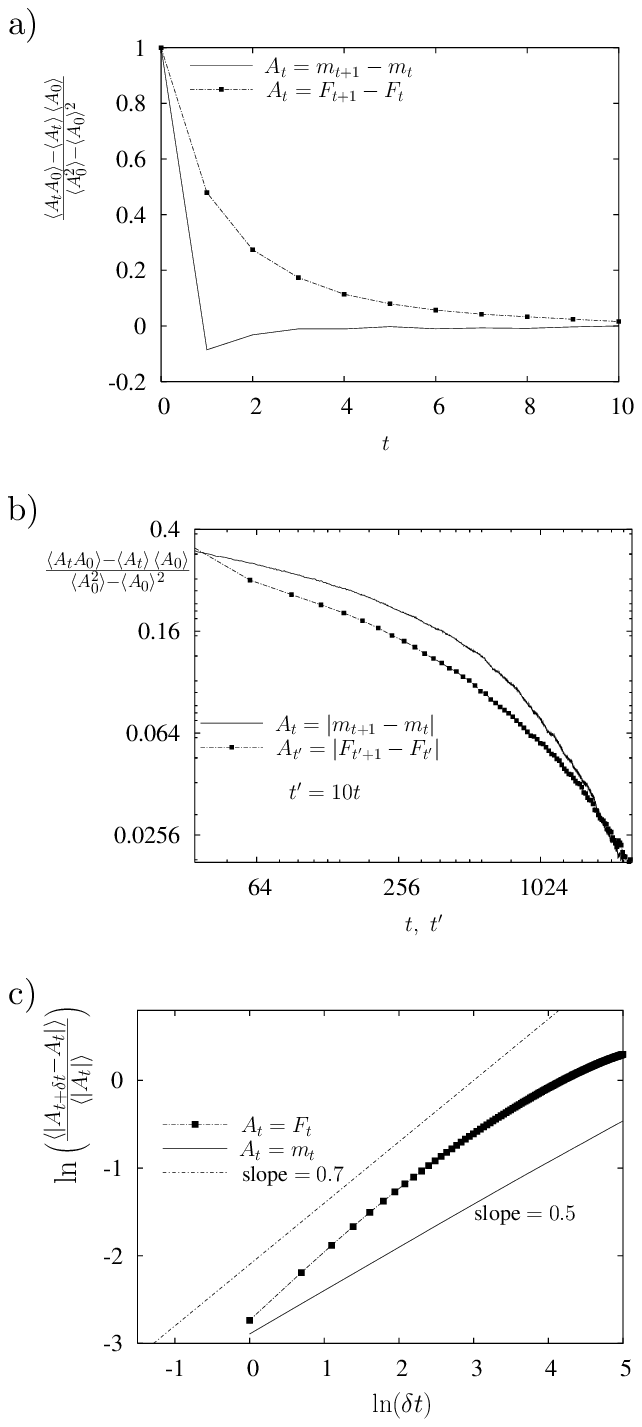, width=8.0cm}
\caption{
Comparison for BKF and our model, as in Fig.\ref{fIG7}, for autocorrelations.
a)~difference between correlations and anticorrelations as a key distinction.
b)~volatility autocorrelations with the rescaling $t'=10\, t$ to emphasize
the similarity of models. However, contrary to the expectations in Ref.~\cite{Yamano2002},
our fits do not confirmed hypothesis about the power-law volatility autocorrelations 
in the spin market models. c)~principal difference between diffusive~(the curve 
with the slope $0.5$ corresponds to BKF) 
and super-diffusive~(slope $0.7$) regime.}
\label{fIG8}       
\end{center}
\end{figure}

\subsection{The comparison with other spin models of financial markets}

The spin model presented in previous sections displays several interesting 
statistical properties that are particularly relevant for stochastic financial 
market dynamics. One of our challenges is to examine cross-links between 
our feedback model and BKF model. In BKF model the local spin behavior is given by
the competition between the local and non-local interactions involved in 
local field $h_{t,i} = 2\,( \sum_{\rm nn} S_{j,t}  -  \kappa S_{i,t} | m_t | )$,
where $\kappa$ is proportionality factor of feedback instant magnetization.  
The probability of $\pm\frac{1}{2}$ lattice spin states 
is given by $1/[1+\exp(\mp 2{\tilde\beta} h_i)]$, where ${\tilde\beta}$ 
is the inverse constant temperature. The difference $m_{t+1}-m_t$ 
is interpreted as a logarithm of the price return. 
Within the BKF model the instant price is proportional 
to $\exp (\lambda m_t)$ term, where $\lambda$ 
is related to the ratio of the 
interacting and fundamental traders. 

In further we assume 
the equivalence of 
$\exp(\lambda m_t)$ and 
$\exp(\lambda F_t)$ terms considered 
for different models. 
The interpretation of 
$\exp(\lambda F_t)$ as a price term 
is in accord with requirement that price increases 
when the competition is weakened. 
The crucial items emphasizing the 
common aspects of models are listed in table~\ref{Tabbb1} 
and illustrated in Figs.~\ref{fIG7} and~\ref{fIG8}. 

We see that both models yield  L\'evy distributions with 
the nearly common 
index $\alpha_1\sim 3.3$ related to cumulative distributions 
with power law slope $-2.3=-\alpha_1+1$ that is consistent with the
empirical data~\cite{Gopikrishnan2000}. The bowl-shaped anomaly at small 
$|F_{t+1}-F_t|$ is probably due to $\mbox{sign}(F_t)$ sharpness. 
The remarkable feature of this comparison 
is a similarity of the exponents obtained 
for pdf's of the bull (bear) market durations. In addition, the similarity 
of temporal aspects can be seen in coincidence of stationary conditional 
probabilities of bull-bear switching listed in the Table~\ref{Tabbb1}. 
We see that $F_t$ feedback model 
acquires the exponential form typical for 
the positively correlated quantities, whereas the 
magnetization variations predicted by BKF are anticorrelated. 
Clearly, this difference can be explained providing energy cumulants
temporal filtering. The simulation of BKF model reveals 
some gap between Hurst exponents~\cite{Kim2004}. 
Contrary to the standard diffusive~$\delta t^{0.5}$ behaviour which
is typical for $\langle | m_{t+\delta t}-m_t | \rangle$,  
the superdiffusive short-time regime 
$\langle |F_{t+\delta t}-F_t| \rangle 
\sim {\delta t}^{0.7}$ has been identified 
for $F_t$ feedback model. 

\begin{table}
\begin{center}
\caption{
The comparison of the statistical characteristics of models of financial
market for $L=90$. 
The simulation results obtained 
for parameters  $\eta=0.01$ and $\Delta=10^{-5}$. 
The BKF parameters $\kappa=20$, ${\tilde\beta}=2$ 
are chosen to attain the intermittent regime. 
The positive $F_t$ (also $m_t>0$) is the signature of 
so called bull market regime, 
while $F_t<0$ ($m_t<0$) is interpreted as a 
bear market regime.}
\label{Tabbb1}
\begin{tabular}{|l||l|l|l|}
\hline
                              &  BKF model             &  model:   $\mbox{sign}(F_t)$                      
& 
\\
\hline 
\hline
price term                    &    $\propto \exp(\lambda m_t)\,,\,$ 
                              &    $\propto \exp( \lambda F_t) $   
&\\
logarithmic returns  &   $ m_{t+1} - m_t $  & $ F_{t+1} - F_t$        
&\\
\hline
pdf of log. returns        &    leptocurtic                       &  leptocurtic
& Fig.\ref{fIG7}a)\\
pdf of bull (bear) durations   &  slope$=-1.3$                             &  slope$=-1.2$ 
& Fig.\ref{fIG7}c)\\
\hline
autocorrelations     &    anticorrelated      &  $\exp(-t/1.6)$ 
& Fig.\ref{fIG8}a)\\
of logarithmic returns     & at $t=1$ & decay &  \\
\hline
volatility of log. returns    &    $|m_{t+1}-m_{t}|$                 &  $|F_{t+1}-F_t|$
&\\
pdf of volat. log. returns    &    $\alpha_1= 3.3$                  &  $\alpha_1= 3.3$                
& Fig.\ref{fIG7}b)\\                           
volat. autocorrelations       &    non-universal                     &  non-universal                   
& Fig.\ref{fIG8}b)\\
diffusion of $F_t, m_t$       &    diffusion $\delta t^{0.5}$        &
\mbox{\small super-diffusive}   $\delta t^{0.7}$
& Fig.\ref{fIG8}c)\\
\hline conditional probabilities 
&  &  &     
\\
$ \pi(\mbox{\scriptsize bull}| \mbox{\scriptsize bull})$ \,\,\,   
$ \pi(\mbox{\scriptsize bull}| \mbox{\scriptsize bear})$   
&  0.978 \,\,\,  0.022   &  0.974 \,\,\,    0.026 &  
\\ 
$ \pi(\mbox{\scriptsize bear} | \mbox{\scriptsize bull}) $ \,\,\,   
$ \pi(\mbox{\scriptsize bear} | \mbox{\scriptsize bear})$   
&  0.022 \,\,\,  0.978   &  0.026 \,\,\,    0.974 &     
\\
\hline
\end{tabular}
\end{center}
\end{table}

\subsection{The application to inter-stock correlations}

In the subsection certain diverse application of our model is presented.
As one sees, the MC model may be used to understand the basic features 
of the mechanism of regulation of bilateral stock cross-correlations. 
The role of central institutions, modeled via the feedback, is to perform 
large-scale eligible decisions that could enhance portfolio efficiency. 
These contribute to $F_t$ comprising a summary of the stock prices. 
We assume that different stocks are distinguished by lattice 
position and sign of dominant trend at given time. 
And for now the correspondence between spins and stocks follows 
a simple majority rule:  
$S_i=\frac{1}{2}$~($S_i=-\frac{1}{2})$ when the {\em sell}~(buy) 
market {\em orders} prevail during elementary time lag~(1~MC step/site),
respectively. The cross-correlations among directly regulated - four 
"nearest" stocks are related to the mean exchange energy term 
$\langle S_i S_{j\in {\rm nn}(i)} \rangle$ 
optimized by $F_t$ to be extremely susceptible 
to regulation. The long-range cross-correlations 
stem from indirect regulation of the portfolio composition near 
to criticality.

\begin{figure}
\begin{center}
\epsfig{file=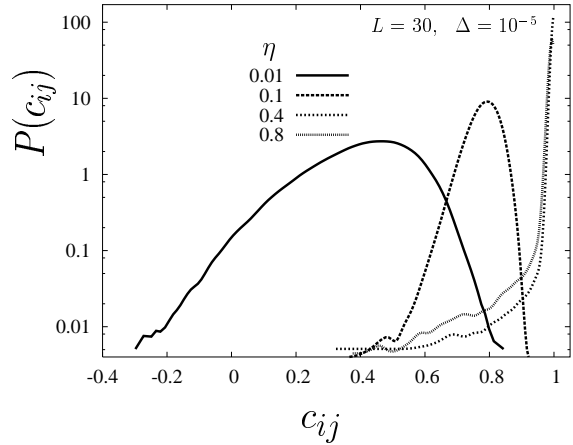,width=8.0cm}
\end{center}
\caption{
Cross-correlations distribution obtained for different $\eta$~[see Eq.(\ref{Cross1})].
The anticorrelations appear for $\eta=0.01$. In that case the distribution can be 
approximated by double-Gaussian $P(c)=\sum_{k=0,1} 
\,w_k$ $\exp \big[ - \big( \, c - a_k\,\big)^2/(2 \sigma_k^2)\,\big]$ 
including the weights
$w_0=1.432$, $w_1=1.951$ and dispersions $\sigma_0=0.146$, $\sigma_1=0.096$
and $a_0= 0.337$, $a_1=0.493$.}
\label{fIG9} 
\end{figure}

The presence of correlations~(and anticorrelations) between 
pairs of stocks play a key role in the theory of
selecting of the most efficient portfolio composition 
of the financial goods~\cite{Mantegna99}. 
For discussed spin model the properties of the pairs 
of stocks can be characterized by 
the cross-correlation coefficient 
\begin{equation}
c_{ij,t} =  
\frac{4}{n_{\rm av}}\,\sum_{t'=t}^{n_{\rm av}+t} S_{i,t'} S_{j,t'}\,,
\qquad i,j \in \{1,2,\ldots, L^2\,\}
\label{Cross1}
\end{equation} 
ranging from $-1$ to $1$. 
Here $i,j$ denote the 
pair of randomly chosen sites. 
In the simulations the temporal averages 
have been calculated for~$n_{\rm av}=10^3$ 
steps and accumulated in pdf's 
depicted in~Fig.\ref{fIG9}. 
We see that the pair correlations 
are dominantly positive 
except a negative tail that agrees qualitatively with the 
empirical studies~\cite{Mantegna99}.
For $\eta$ larger than $\sim 0.1$ 
the negative tail vanishes and pdf 
of $c_{ij,t}$ is shifted towards $1$. 
Clearly, such "coherent" regime is 
typical for inefficient 
portfolio structure.

\section{Conclusion}
In conclusion, we have presented application of adaptivity concept to critical phenomena.  
Despite of the principal difficulties, insufficient understanding 
of the systematic errors which arise proceeding from non-equilibrium 
to equilibrium, we expect that work might inspire numerous applications 
to other inverse statistical problems where constrains are written 
in terms of running averages. 

We have demonstrated that MC model can be reinterpreted in macro-economic terms. 
The simulations indicated spontaneous emergence of 
the power-law distribution for feedback regulatory signals which is 
universal feature of the stationary  non-equilibrium regime.  
Surprisingly, we found that $F_{t+1}-F_t$ statistics contradicts to 
our previous expectations based on the finding
$\mbox{C}_t^{(2)}\sim t^{-q'}$,      
which does not imply power-law fall-off of the autocorrelations of 
$|F_{t+1}-F_t|$. 
The influence of the memory depth related to parameter $\eta$ 
has also been investigated extensively. The critical 
point related to the most efficient portfolio structure, is an abstract 
of ideal situation when $\eta\rightarrow 0^{+}$, which is  
unattainable in a real finite-memory conditions. 
Otherwise, we have found that the portfolio efficiency is weakened
via short-time memory for $\eta>0.5$.
The comparative study of two spin models have 
confirmed that presented MC dynamics based on the specific heat feedback 
offers an alternative description of financial markets. 
Despite of outstanding similarities, the comparison 
of exponents indicate non-trivial relationship between 
the universality classes of studied models.

\noindent {\bf Acknowledgment}

The authors would like to express their thanks to Slovak Grant agency 
VEGA (grant no.1/2009/05) and agency APVT-51-052702 for financial support.


\begin{thebibliography}{999}

\bibitem{Melby2000}
P.~Melby, J.~Kaidel, N.~Weber, A.~H\"ubler,  Phys.Rev.Lett. {\bf 84} (2000) 5991.
\bibitem{YoussHolyst2000}
M.~Youssefmir, B.A.~Huberman, 
T.~Hogg, Computational Economics {\bf 12} (1998) 97;

 J.A.~Holyst, K.Urbanowicz, Physica A {\bf 287} (2000) 587.  
\bibitem{Bleich1996} 
M.E.~Bleich, J.E.S.~Socolar, Phys.Rev.E  {\bf 57} (1996) R16.
\bibitem{HorvathGmitra2004}
 D.~Horv\'ath, M.~Gmitra, Z.~Kuscsik, Czech.J.Phys. {\bf 54} (2004) 921;
  
 D.~Horv\'ath, M.~Gmitra, Int.J.Mod.Phys.C. {\bf 15} (2004) 1269.

 \bibitem{Binder1988}
 K.~Binder, D.W.~Heermann, Monte Carlo Simulation in Statistical Physics,  
 Springer, Berlin, 1998.
 \bibitem{Principe2000}
  J.C.~Principe, N.R.~Euliano,    
  W.C.~Lefebvre, Neural and adaptive systems: 
  Fundamentals through simulations, 
  John Wiley \& Sons, New York, 2000.
 \bibitem{Metropolis}
  N.~Metropolis, 
  A.W.~Rosenbluth, 
  M.N.~Rosenbluth, 
  A.H.~Teller, J.Chem.Phys. {\bf 21} (1953) 1087.
 \bibitem{Buonsante2002}
 P.~Buonsante, R.~Burioni, D.~Cassi, A.~Vezzani, Phys.Rev.E {\bf 66} (2002) 36121.  
\bibitem{Garrido87}
 P.~L.~Garrido, A.~Labarta, J.~Marro, J. Stat. Phys. {\bf 49} (1987) 551;

 J.~M.~Gonzales-Miranda, P.L.~Garrido, 
 J.~Marro, J.~L.~Lebowitz, 
 Phys.Rev.Lett. {\bf 59} (1987) 1934;

 K.~Basler, Z.~R\'acz, Phys.Rev.Lett. {\bf 73} (1994) 1320.   
\bibitem{Tomita2001} 
  Y.~Tomita, Y.~Okabe, Phys. Rev. Lett. {\bf 86} (2001) 572.
\bibitem{Angelini2003} 
 L.~Angelini, Physics Letters A {\bf 307} (2003) 41.
 \bibitem{ExactSolOnsager1944}
  L.~Onsager, Phys. Rev. {\bf 65} (1944) 117;
  
  R.J.~Baxter, Exactly Solved
  Models in Statistical
  Mechanics, Academic Press, London, 1982.
\bibitem{LandauBinder2000}
D.P.Landau, K.Binder, A guide to Monte Carlo simulations 
in statistical physics, Cambridge University Press, 
Cambridge, 2000.
 \bibitem{Bak88}
   P.~Bak, C.~Tang, K.~Wiesenfeld, Phys.Rev.A {\bf 38} (1988) 364;

   P.~Bak, C.~Tang, K.~Wiesenfeld, Phys.Rev.Lett. {\bf 59} (1987) 381;

   C.~Tang, P.~Bak, Phys.Rev.Lett. {\bf 60} (1988) 2347. 
\bibitem{Drossel92}
  B.~Drossel, F.~Schwabl,  Phys.Rev.Lett. {\bf 69} (1992) 1629.
\bibitem{Alstrom94}
   P.~Alstrom, J.~Le${\tilde {\rm a}}$o, 
   Phys.Rev.Lett. {\bf 49} (1994) R2507.
\bibitem{Kadanoff}
  L.P.~Kadanoff, Physics Today (March 1991) p.~9
 \bibitem{Sornette}
 D.~Sornete, J. Phys. I  France {\bf 2} (1992) 2065;

 D.~Sornette, A.~Johansen, I.~Dornic, 
 J.Phys. I France {\bf 5} (1995) 325.

\bibitem{Fortuin72}
   C.M.~Fortuin, P.W.~Kasteleyn, Physica {\bf 57} (1972) 536.
  \bibitem{Wolff89}
  U.~Wolff, Phys. Rev. Lett. {\bf 62} (1989) 361. 
\bibitem{Mouritsen92}
  J.V.~Andersen, O.G.~Mouritsen, Phys.Rev.A {\bf 45} (1992) R5331.

 \bibitem{Mantegna99} 
  R.N.~Mantegna and 
  H.E.~Stanley, Introduction 
  to Econophysics: 
  Correlations and Complexity in Finance, 
  Cambridge University Press, Cambridge, 1999.


\bibitem{Langton1990}
 Ch.G.Langton, 
 "Life at the edge of chaos", p.41 in Artificial life II, 
 Proceedings of the workshop 
 on artificial 
 life held february 1990 in Santa Fe, 
 New Mexico, Proceedings Vol.X, Addison-Wesley, 1990.
 \bibitem{Ponzi2000}  A.~Ponzi, Y.~Aizawa,  Physica A {\bf 287} (2000) 507;

  T.~Kaizoji, Physica A {\bf 287} (2000) 493;

  A.~Krawiecki, J.A.~Holyst, D.~Helbing, 
  Phys. Rev. Letters {\bf 89} (2002) 158701;

  \bibitem{Bornholdt2001}
  S.~Bornholdt, Int.~J.~Mod.~Phys.~C {\bf 12} (2001) 667;

  T.~Kaizoji, S.~Bornholdt, Y.~Fujiwara, Physica A {\bf 316} (2002) 441.

  M.~Badshah, R.~Boyer, T.~Theodosopoulos, math.PR/0501244, 
  "{\it Statistical properties of the phase transitions 
  in a spin model for market microstructure}", 
  January 2005.
   
  \bibitem{Challet1997}
   D.~Challet, A.~Chessa, M.~Marsili, Y.-C.~Zhang, 
   Quantitative Finance {\bf 1} (2001) 168; 

   D.~Challet, Y.-C.~Zhang, 
   Physica A {\bf 246} (1997) 407;
  
  D.~Challet, Y.-C.~Zhang, 
  Physica A {\bf 256} (1998) 514. 

\bibitem{Wrigley1992} 
N.~Wrigley, Environment and Planning A {\bf 24} (1992) 727. 

\bibitem{Yamano2002} T.~Yamano, Int.J.Mod. Phys. C, {\bf 13} (2002) 89.

\bibitem{Gopikrishnan2000}
P.~Gopikrishnan, 
V.~Plerou, Y.~Liu, L.A.N.~Amaral, X.~Gabaux, H.E.~Stanley, Physica A, 
{\bf 287} (2000) 362.

\bibitem{Kim2004} 
K.~Kim, S.M.~Yoon, 
Physica A {\bf 344} (2004) 272.

\end{thebibliography}
\end{document}